\documentstyle[preprint,revtex]{aps}
\begin{document}
\draft

\begin{title}
A Simple Model for Coupled Magnetic and Quadrupolar Instabilities in
Uranium Heavy-Fermion Materials
\end{title}

\author{V. L. L\'{\i}bero}
\begin{instit}
Departamento de F\'{\i}sica e Ci\^encia dos Materiais, Instituto de F\'{\i}sica
e Qu\'{\i}mica de S\~ao Carlos,
Universidade de S\~ao Paulo, 13560 S\~ao Carlos, S\~ao Paulo, Brazil
\end{instit}

\author{ D. L. Cox}
\begin{instit}
Department of Physics, Ohio State University, Columbus, Ohio 43210 \\
\end{instit}

\begin{abstract}
We present a mean-field calculation of the phase diagram of a simple
model of localized moments, in the hexagonal
uranium heavy-fermion compounds.
The model considers a non-Kramers quadrupolar doublet ground state
magnetically coupled with a singlet excited-state, favoring in-plane
van-Vleck magnetism, as has been conjectured for UPt$_3$. The Hamiltonian
which defines the model is Heisenberg like in both, magnetic and
quadrupolar moments.  No Kondo effect physics is included in the calculations.
 Among our main results are: (i) for zero intersite quadrupolar
coupling,  the magnetic order is achieved by a first order transition above a
critical
intersite magnetic coupling value which becomes second order
at higher coupling strengths. (ii) for
finite intersite quadrupolar coupling, at temperatures below a second
order quadrupolar ordering transition,
the minimal magnetic coupling value is
increased, but (a) the magnetic ordering temperature is enhanced above this
value, and (b) the ordering of first and second order transitions in the phase
diagram is reversed.  By considering the general structure of the
Ginsburg-Landau free
energy,  we argue that the Kondo effect will not modify the shape of
the phase diagram, but will modify the quantitative values at which transitions
occur.
\end{abstract}


\narrowtext
\section{INTRODUCTION}
\label{sec:intro}

The unusual commensurate, weak moment
magnetic order present in heavy electron superconductors such as
UPt$_3$ [and its related alloys U(Pt$_{1-x}$,Pd$_x$)$_3$ and
(U$_{1-x}$,Th$_x$)Pt$_3$]
and URu$_2$Si$_2$ has so far resisted explanation in any
simple terms.  It is difficult to believe the small moment order
represents itinerant spin density wave formation, since, for example,
UPt$_3$ displays no appropriate nesting features on the Fermi
surface at the appropriate anti-ferromagnetic wave vector.  It is
difficult to account for the order in terms of frustration effects
since the ordered structures are remarkably simple.  In the case
of URu$_2$Si$_2$, moreover, there is evidence \cite{Broholm} that the magnetic
order parameter is not the `primary' one, in the sense that
despite the large entropy involved in the 17.5K magnetic
transition (the entropy involved is larger than in the 1.2K
superconducting transition) the effective moment is too small to
be consistent with the size of the ordered moment.  Indeed, it has
been suggested \cite{Ramirez} that some other ``hidden'' order drives the
transition (such as quadrupoles).

Recently results coming from different directions altogether make
it prudent to consider quadrupolar
(and octupolar)
ordering possibilities in these materials:  it has been noted that
the quadrupolar Kondo model believed to be previously restricted
to cubic uranium based materials is in fact possible for hexagonal
(UPt$_3$) and tetragonal (URu$_2$Si$_2$) materials as well \cite{quadruKondo}.
In this model the heavy fermions arise from local
uranium quadrupole moments associated with ground crystal field
doublets quenched by (predominantly) the orbital motion
of itinerant electrons.  Thus, the low lying degrees of freedom
may be essentially quadrupolar.

The relevance of the quadrupolar
Kondo model to this class of materials has been put on firmer
footing by the discovery of the cubic material
Y$_{1-x}$U$_x$Pd$_3$, which appears to display a dilute limit
quadrupolar Kondo effect \cite{YUPd}.   Moreover, all known heavy fermion
superconductors possess the appropriate symmetry groups to yield a
quadrupolar Kondo effect or magnetic two-channel Kondo effect
(CeCu$_2$Si$_2$), a point bolstered by the recent discovery of two
hexagonal heavy-electron superconductors (UPd$_2$Al$_3$ and
UNi$_2$Al$_3$) which also possess magnetic order \cite{Geibel}.

In the quadrupolar Kondo picture, a context for
weak moment magnetic order can arise
naturally from two distinct sources: (i) for cubic symmetry, all
magnetic character is in excited states, and for hexagonal and
tetragonal symmetry, in plane magnetic character is in excited
states. Thus, the magnetic response is van Vleck in character
yielding a built in stability against magnetic ordering and a
region of weak moment order. (ii) For hexagonal and tetragonal
symmetry, the $c$-axis character of the local uranium doublets is
weakly magnetic, but primarily octupolar.

In this paper, we shall focus our attention on (i), with the
particular goal of examining how quadrupolar order can induce or at least
enhance magnetic ordering tendencies.  While it is well known that
quadrupolar order will always be induced as a secondary order
parameter at a magnetic transition, it seems to be less well
known, at least in the heavy fermion field, for magnetic order to
be induced or enhanced by the coupling to quadrupolar degrees of
freedom.  Such physics is known
in UO$_2$, for example, where the antiferromagnetic phase
transition is first order due to a coupling between magnetic and
quadrupolar order parameters \cite{Allen}.

We have performed mean field calculations on a simple model of
potential relevance to the hexagonal uranium heavy fermion
superconductors.  In this model, each uranium ion possesses a
doublet quadrupolar
ground state magnetically coupled to an excited singlet
state.  The uranium ions are placed on a lattice with the
structure of UPt$_3$ (though the point symmetry is taken to be
$D_{6h}$ rather than $D_{3h}$).
We then turn on nearest neighbor in-plane and out-of-plane
quadrupolar and magnetic coupling
and study the resulting phase diagram. We find a rich structure,
ranging from van Vleck magnetism with a tricritical point and
critical exchange strength to
alternately suppressed and enhanced magnetic order below a second
order quadrupolar transition.  We discuss the phase diagram using
both the full mean field calculations and Landau theory in regions
of validity.

While the model does not produce a full explanation of the
ordering in bulk UPt$_3$ (for which
the specific heat shows no transition and the neutron elastic scattering
peaks are not true resolution limited Bragg peaks) it may be of
relevance to the Pd and Th doped alloys and to the new hexagonal
superconductors.  In any case, it represents the beginning of an
alternative view of collective phenomena in the heavy fermion
materials which may ultimately provide a unifying view of the
heavy fermion state, superconductivity, and magnetism.

The outline  of the paper is as follows: Section II provides an
overview of the model.  Section III develops the mean field
approximation to this model.  Section IV discusses our results in
terms of Landau theory, paying special attention to the role of
the orientation dependent coupling between quadrupolar and
magnetic order parameters.  Finally, in Section V we conclude and
point out explicit experimentally relevant features of our model.

\section{MODEL}
\label{sec:model}

        In the section II-A we write a single-site Hamiltonian, coupling the
quadrupolar moment of the ground state of the uranium-ion with the
electric-field
gradient and the magnetic moment with an applied magnetic field.
In the next section we introduce intersite coupling of the XY-model form
for both quadrupolar and magnetic degrees of freedom.
Within the mean-field approximation that Hamiltonian is equivalent to
that of section II-A, but with coupling coefficients which depend
self-consistently on the mean magnetic and quadrupolar moments.

\subsection{Single-Site Hamiltonian}
\label{subsec:singsit}

	Having in mind the motivations given in the Introduction, we define
a single-site model in which the uranium ion has a non-Kramers quadrupolar
doublet ground state
magnetically coupled only with an excited state. Such a crystal-field scheme
may arise for a U$^{4+}$, $5f^2$, $J=4$ at a site of hexagonal
symmetry. Fig.~\ref{3levels} defines the
level structure of the model, where the doublet $|E_{\pm}\rangle$ and
the singlet $|B\rangle$ are defined by
\begin{equation}
	|E_{\pm}\rangle = u|\pm 4\rangle + v|\mp 2\rangle \;,
\end{equation}
and
\begin{equation}
        |B\rangle = \frac{|3\rangle + |-3\rangle}{2}  \;,
\end{equation}
where $|m\rangle$ are the eigenstates of the z-component of the total angular
momentum $J$. u and v are constants. The relevant matrix elements for the
model we are defining are
\begin{equation}
	\langle B| J_{\pm} |E_{\mp}\rangle = 2u+\sqrt{7}v \equiv a \;,
\label{eq:a} \end{equation}
and
\begin{equation}
        \langle E_{\pm}| J_{\pm}^2 |E_{\mp}\rangle = 2 \sqrt{7}v \equiv b \;.
\label{eq:b} \end{equation}
Since $u^2+v^2=1$, the constants a and b defined here are not independent.

	The other $J=4$ multiplet levels are not considered,
so the model will be restricted to temperatures less than or equal to
the first excitation energy $\Delta$. It is
clear from the level structure considered that this model favor in-plane
van Vleck magnetism, and c-axis Pauli magnetism, as has been conjectured
for UPt$_3$ \cite{quadruKondo}.  The motivation for focussing on the doublet
level is that it is the
only degenerate level in hexagonal symmetry for U$^{4+}$ ions, and we desire
the internal degrees of freedom associated with the doublet level to allow for
the Kondo
effect.  The formation of the many body Abriksov-Suhl-Kondo resonance is the
basic paradigm for the source of heavy fermions in these materials.

	 The doublet ground state corresponds to the $E_2$ representation
of the point-group $D_{6h}$, where the eigenfunctions transform like
$x^2-y^2$ and $xy$. This doublet has magnetic character along the c-axis,
but only in-plane quadrupolar tensors may couple the doublet levels
together. This implies we can have a quadrupolar Kondo effect.
Using the Stevens operator representation \cite{sor},
the corresponding quadrupolar operators are (proportional to) $Q_{xx}=-Q_{yy}=
J_{x}^2-J_{y}^2$ and $Q_{xy}=Q_{yx}=(J_xJ_y+J_yJ_x)/2$. The magnetic
moment operator is $\vec{M}=g_J \mu_B \vec{J}$, where $g_J=4/5$ and
$\mu_B$ is the Bohr magneton. The doublet is connected by $J_x$, and $J_y$ to
the excited  $|B\rangle$ singlet. [Note that while the actual point group at U
sites in
UPt$_3$ is $D_{3h}$, the allowed crystal field spectrum is essentially
identical to that of $D_{6h}$.]

	A single-site Hamiltonian $H$ is defined coupling the quadrupolar
operators with an electric-field gradient $\nabla \vec{E}$ and
coupling the magnetic
moment with an external magnetic-field $\vec{H}$, at the uranium site:

\begin{equation}
H=H_{0}-\vec{H} \cdot \vec{M} -
\sum_{\alpha,\beta} (\nabla_{\alpha} E_{\beta})
\hat{Q}_{\beta \alpha} \;,
\label{eq:grad}
\end{equation}
where $H_0$ is the Hamiltonian for the uranium ion in the presence of the
crystal field only (see Fig.~\ref{3levels}):
\begin{equation}
H_0  = \Delta |B\rangle \langle B| \;.
\label{eq:h0}
\end{equation}
We are measuring energies from the doublet ground-state $|E_{\pm}\rangle$.

	For in-plane $\vec{H}$ and $\vec{E}$, the Hamiltonian
Eq.~(\ref{eq:grad}) is written as
\begin{equation}
H = H_0 + AJ_+ + BJ_+^2 + h.c. \;,
\label{eq:single} \end{equation}
where
\begin{equation}
A = -\frac{g_J\mu_B}{2} (H_x - i H_y)
\label{eq:A1} \end{equation}
and
\begin{equation}
B = -\frac{\lambda}{2} (\partial_{xx}E - \partial_{yy}E- 2i\partial_{xy}E) \;.
\label{eq:B1} \end{equation}

	The diagonalization of this Hamiltonian shows that there is a
crossing of levels depending upon $\vec{H}$ and $\vec{E}$ and an
interplay between the field gradient and magnetic field which
depends crucially on their orientation with respect to one another.
In the model defined in the next section, the coefficients $A$ and $B$ are
interpreted as molecular fields acting on the uranium sites.
Since those molecular
fields depend on the magnetic and quadrupolar moments, a self-consistent
calculation is required to obtain these moments
at each temperature.

\subsection{The Intersite-coupling Hamiltonian}
\label{subsec:intersite}

	In order to determine the molecular fields $\vec{H}$ and $\vec{E}$
present in Eq.~(\ref{eq:grad}) we have used a model Hamiltonian
which is Heisenberg like
in the magnetic moments as well in the quadrupolar moments, defined by

\begin{eqnarray}
H &=& H_{0} - \sum_{<i,j>} ( I^{ab}_{m} \vec{M_{i}} \cdot \vec{M_{j}} +
I^{c}_{m} \vec{M_{i}} \cdot \vec{M_{j}}\nonumber\\
&+& I^{ab}_{q} Tr \hat{Q_{i}} \cdot \hat{Q_{j}} +
I^{c}_{q} Tr \hat{Q_{i}} \cdot \hat{Q_{j}} ) \;,
\label{eq:H}
\end{eqnarray}
where $I^{ab}_{m}$ $(I^{c}_{m})$ is the coupling between the
nearest-neighbour magnetic moments in the basal plane (c-axis direction) of
the hexagonal lattice; $I^{ab}_q$ $(I^{c}_q)$
has the same meaning for the quadrupolar moments. $Tr$ means trace over
the product of the quadrupolar tensors.

	We are interested in studying the antiferromagnetic order of
the model defined by Eq.~(\ref{eq:H}), so that the couplings $I^{ab}_{m}$
and $I^{c}_{m}$ are negative. Only ferroquadrupolar coupling
$J_q$ has been considered. We have examined the effect of a coupling of the
form $Tr \vec{M_i} \cdot \hat{Q_j} \cdot \vec{M_k}$ in the Hamiltonian
Eq.~(\ref{eq:H}). It does not change qualitatively the mean-field phase
diagram of the model, since the couplings already present in Eq.~(\ref{eq:H})
give rise to a term proportional to $M^2Q$ in the free energy
(see Eq.~(\ref{eq:fexp})). Hence we will drop this coupling from
the Hamiltonian.
The calculations are restricted to the case when $\vec{M}$ lies in
$x$-direction (that
is the case for UPt$_3$, as showed by neutron scattering data \cite{nscat}),
although it is possible to consider the general case.

\section{Mean-Field Theory}
\label{sec:mean}

\subsection{Mean-Field Hamiltonian}
\label{subsec:1site}

	The mean field approach consists in replacing the operators in
the Hamiltonian by their mean values plus the respective fluctuations,
and keeping the fluctuations only up to first order. Denoting the
mean value of the magnetic and quadrupolar moment by
$\bar{M}\equiv g_J\mu_Ba\bar{m}$
and $\bar{Q}\equiv b\bar{q}$, respectively, where the matrix elements
$a$ and $b$ are defined in Eq.~(\ref{eq:a}) and Eq.~(\ref{eq:b}),
the mean field applied to Eq.~(\ref{eq:H}) results in

\begin{equation}
\frac{H}{\Delta} = \frac{H_0}{\Delta} - N(A\bar{m}+B\bar{q}) +
			2\sum_{i} (A m^{x}_i + B q^{xx}_i) \;,
\label{eq:Hamil} \end{equation}
where $N$ is the number of sites of the hexagonal lattice,
\begin{equation}
A=\frac{(I^{ab}_M +I^{c}_M)a^2}{\Delta}\bar{m} \equiv J_m \bar{m} \;,
\label{eq:A} \end{equation}
and
\begin{equation}
B=-6\frac{(I^{ab}_Q + I^{c}_Q )b^2}{\Delta}\bar{q} \equiv -6 J_q \bar{q} \;.
\label{eq:B} \end{equation}
The last equalities, Eq.~(\ref{eq:A}) and
(\ref{eq:B}), define the dimensionless magnetic coupling $J_m$ and the
dimensionless quadrupolar coupling $J_q$, respectively. We are
using deliberately the same notation $A$ and $B$ here as in
Eq.~(\ref{eq:A1}) and (\ref{eq:B1}) since in both cases the meaning of
those parameters are the same, although in Eq.~(\ref{eq:A}) and
(\ref{eq:B}) $A$ and $B$ depend on the order parameters $\bar{m}$
and $\bar{q}$.

	The site-independent terms of $H$ are called the condensed energy;
the other terms define a sum over an effective one-site Hamiltonian $H_{i}$,
which can be written in terms of the components of the total angular
momentum as
\begin{equation}
H_i = H_{0i} + A(J_{i+} + J_{i-}) + B(J^{2}_{i+} +J^{2}_{i-}) \;.
\end{equation}

 	Diagonalizing $H_i$ in the subspace defined by the electronic
levels shown in Fig.~(\ref{3levels}) we find the following expression
for the free-energy per site and in units of $\Delta$:

\begin{equation}
f = -J_m\bar{m}^2+6J_q\bar{q}^2-t \ln( e^{6\beta J_q \bar{q}^2} +
	2 e^{-\beta (1-6J_q \bar{q}^2)} \cosh{\beta C} ) +
	t \ln( 2+e^{-\beta} ) \;,
\label{eq:free} \end{equation}
where $\beta = 1/t$ is the inverse of the dimensionless temperature
$t \equiv T/\Delta$. $C$ is given by
\begin{equation}
C=\sqrt{ (\frac{1+6J_q \bar{q}}{2})^{2}+2 J_m^2\bar{m}^2 } \;.
\end{equation}

	From the free-energy, we obtain the following coupled equations for the
order parameters $\bar{m}$ and $\bar{q}$ :

\begin{equation}
\bar{m} = -2\frac{J_m \bar{m}}{C} \frac{\sinh{\beta C}}{ exp[\beta (1-
		18J_q \bar{q})/2] + 2\cosh{\beta C} } \;,
\label{eq:mbar} \end{equation}
and
\begin{equation}
\bar{q} = \frac{1}{2} \frac{ \cosh{\beta C}+
	   (1+6J_q \bar{q})/(2C) \sinh{\beta C}-
	     exp[\beta (1-18J_q \bar{q})/2] }{exp[\beta (1-18J_q \bar{q})/2]+
		2\cosh{\beta C} } \;.
\label{eq:qbar} \end{equation}

	Before solving equations (\ref{eq:mbar}) and
(\ref{eq:qbar}) numerically, we describe a few limits where it
is possible to study these equations analytically. There is no solution
other than the trivial $\bar{m}=0$ and $\bar{q}=0$ if $J_m<-1/2$ and
$J_q=0$. A pure magnetic phase ($\bar{m} \ne 0$, $\bar{q}=0$) does not
exist, but a pure quadrupolar phase ($\bar{m} = 0$, $\bar{q} \ne 0$)
exists for $|J_m| < (1+3J_q)/2$ and $J_q \ne 0$. In
that phase, for $T << \Delta$ one has $\bar{q} \approx 0.5\tanh{(6\beta J_q
\bar{q})}$, which implies the quadrupolar degrees of freedom are playing
the role of a pseudospin $1/2$. In this regime, a second-order
critical line appears at $t_{cq} \approx 3J_q$.
The other regimes of Equations (\ref{eq:mbar}) and
(\ref{eq:qbar}) are studied numerically. The next section
discusses the phase diagram obtained by studying the stabilities of
the solutions $\bar{m}$ and $\bar{q}$ for different temperatures and
values of the coupling coefficients $J_m$ and $J_q$.

\subsection{Phase Diagrams}
\label{sec:result}

	Fig.~\ref{diagram} shows the phase diagram of the model defined
by Eq.~(\ref{eq:Hamil}). The main
features of this diagram are the enhancement of the critical temperature
and the interchange between first and second-order transitions when
the quadrupolar coupling $J_q$ is introduced.

	We first consider the case when $J_q=0$. Both order parameters,
$\bar{m}$ and $\bar{q}$,
are critical simultaneously and despite the fact that we do not have
quadrupolar coupling in this case, the ground state described
by the Hamiltonian $H$, Eq.~(\ref{eq:Hamil}), has quadrupolar degrees
of freedom and for different temperatures the stability of this ground state
against the intersite magnetic coupling causes the
tricritical point (denoted by TP), separating
the first-order and second-order transitions. The coordinates of this
tricritical point are $J_m=-1.05$ and $t_c=0.300$.
The line of second order is obtained from the
limit $\bar{m}$ and $\bar{q}$ going to zero in Eq.~(\ref{eq:mbar}) and
Eq.~(\ref{eq:qbar}) and is given by: $e^{\beta_c} =
(J_m-1/2)/(J_m+1)$. Near this line $\bar{m}$ follows the usual
$(t_c-t)^{1/2}$ behavior characteristic of the order parameter in
mean-field theory; $\bar{q}$ is proportional to $\bar{m}^2$, which is
characteristic of the secondary order-parameter.
The first-order line extends from the tricritical point to
$J_m=-1/2$; below this point no order is found for either $\bar{m}$ or
$\bar{q}$.

	For non-zero $J_q$ we have simultaneously first-order transition for
both order parameters only above the critical-end-point, denoted CEP in
Fig.~\ref{diagram}. Below this point the transitions are separated in one
line of second-order, purely quadrupolar transition at
$t_{cq}=6J_q/(2+e^{-1/t_{cq}})$, and
another line of first-order purely magnetic transition. This latter line has a
tricritical point, denoted TP, becoming
second-order below this point and finishing at $t=0$ and
$J_m=-(1+3J_q)/2$; below this value of $J_m$ one has only quadrupolar order.
In Table~\ref{coord}
we have listed positions of the points CEP and TP for a few values of
$J_q$. Fig.~\ref{behavior} shows typical behavior of both order-parameters
with temperature, in different parts of the phase diagram.

	Besides changing the values of the critical temperatures and
separating the phases below the CEP, the intersite quadrupolar coupling
has a more dramatic effect on the phase diagram, inverting the nature
of the phases with respect to the TP: the first-order transition occurs
below the TP when $J_q=0$, and exactly
the opposite occurs when $J_q \neq 0$. We believe this complicated effect
of the indirect
magnetic-quadrupolar coupling generated by the Hamiltonian Eq.~(\ref{eq:H})
exists even for a very small quadrupolar coupling. This is confirmed by
Landau's mean field theory, which is discussed in the Section
{}~\ref{sec:landau}.

\subsection{Specific-Heat Jumps}
\label{subsec:latent}

	 Table~\ref{jump} shows numerical calculations of the value of the
specific heat jumps, $\Delta C$, for the second-order magnetic transitions
for different quadrupolar coupling $J_q$.
For zero $J_q$, $\Delta C$ is never less than $6 J/mol-K$. For non-zero
$J_q$, $\Delta C$ is a very sensitive function of $J_m$, giving values of
$\Delta C$ around $2 J/mol-K$ for $J_q=0.05$, and $J_m=-0.6$, which is
the experimental value of $\Delta C$ for the compound U$_{1-x}$Th$_x$Pt$_3$.
For these $J_m$ and $J_q$, and considering that the Neel temperature is equal
to $6 K$ for this compound, the model also gives the value of energy
splitting $\Delta$ of the crystal field as $70 K$,
which is the order of magnitude expected for this material.
Naturally we will have corrections to the specific-heat jump
when including explicitly the interaction  with the
conduction-band electrons. If these corrections are not too large, the
model shows that a very small quadrupolar coupling, about one order of
magnitude smaller than the magnetic coupling, is necessary to obtain
the magnitude of the specific-heat jump.

\section{Landau's Mean-Field Theory}
\label{sec:landau}

	Contrary to the previous section, which treated the
molecular-field-theory free energy
exactly, now we want to use the Landau's mean field theory, which is valid
in the limit of small order parameters, i.e., near a second-order
transition. This theory, although limited,
can describe some of the features of the phase diagrams.

\subsection{Free Energy}
\label{subsec:landaufree}

	The Landau's mean-field theory uses an approximation for
the free energy, valid near the region where the order parameters
$\bar{m}$ and $\bar{q}$ are small.
Expanding the free energy given by  Eq.~(\ref{eq:free}) around
$\bar{m}=0$ and $\bar{q}=0$ we have:

\begin{equation}
f=\alpha_m \bar{m}^2 + \beta_m \bar{m}^4 + \gamma_m \bar{m}^6 +
\alpha_q \bar{q}^2 + \beta_q \bar{q}^4 +
\delta_{mq} \bar{m}^2 \bar{q} + \epsilon_{mq} \bar{m}^2 \bar{q}^2 + ... \;,
\label{eq:fexp} \end{equation}
where the coefficients are functions of the temperature $t$ and of the
couplings $J_m$ and $J_q$. They are listed in Table~\ref{fcoef}.

	For $J_q=0$, only the three first terms on the right hand side of
Eq.~(\ref{eq:fexp}) survive and
the critical temperatures for second-order transitions are obtained
from the equation
$\alpha_m(t_c,J_m)=0$, whose solutions agree with the numerical
results. The  coefficient $\beta_m(t,J_m)$ is negative (positive) for
temperature less (great) than 0.30032, and $\gamma_m$ is positive (negative)
for temperature less (great) than 0.46370,
for all $J_m$. That means we have first (second) order for critical
temperatures less (great) than 0.30032. The point where both coefficients,
$\alpha_m$ and $\beta_m$, vanish
simultaneously determines the tricritical point (TP) and that is at
$t_{TP}=0.30032$ and $J^{TP}_m=-1.07140$, also in agreement
with the phase diagram Fig.~(\ref{diagram}).

	The quadrupolar-coupling $J_q$ complicates the analysis of the
free energy Eq.~(\ref{eq:fexp}).  Now, the second-order magnetic transition
(dashed line in Fig.~\ref{diagram}) has non-zero quadrupolar order-parameter
$\bar{q}=q_0$, so the free energy in fact should be written as
\begin{equation}
f=\alpha'_m(t,J_m,q_0)m^2 + \beta'_m(t,J_m,q_0)m^4 + ... \;,
\end{equation}
where $q_0$ is the equilibrium value of $\bar{q}$ on the critical
line ($\bar{m}=0$),
and is given by $q_0=(-\alpha_q/2\beta_q)^{1/2}$.  The
critical temperature for the magnetic transition now is the solution of
$\alpha'_m(t_c,J_m,q_0) \approx \alpha_m+\gamma q_0+\delta q^2_0 + ... =0$,
which gives $t_c$ close to the numerical results, mainly  near (but at
left of) the tricritical point where $q_0$ is not to big.
To calculate $\beta'_m$ needs some care. Using the expansion (\ref{eq:fexp})
$\beta'_m$ will have divergent terms at $t_{cp}$,
which invalidate its expansion around the tricritical point since this point
occurs always near $t_{cp}$. For $J_q=0$ those terms do not exist and
$\beta'_m=\beta_m$. We have computed $\beta'_m$ through
its definition: $\beta'_m=(\partial^4 f/\partial m^4)/4!$, calculated at
$\bar{m}=0$ and $\bar{q}=q_0$, where $q_0$ is obtained from the numerical
solution of the Eq.~(\ref{eq:mbar}) and Eq.~(\ref{eq:qbar}). $\beta'_m$
showed no divergence and, although $\beta_m$ can be negative, $\beta'_m$
is positive for $t \leq t_{TP}$, which explains why we have a second-order
transition when $J_q \neq 0$ intead of a first-order one when $J_q=0$.
Unfortunately, Landau's mean-field theory does not work well right
above the tricritical point TP since $\bar{m}$ does not go to zero,
and in fact $\bar{m}$
has a big jump at the critical temperature and so $\bar{q}$ has a
discontinuity at this line (see Fig.~\ref{behavior}e). For large $J_m$,
however, the jumps in both order parameters at the critical line become
small, thus the expansion (\ref{eq:fexp}) gives a reasonable description of
the model in this limit.

\section{Summary and Conclusions}

	We have calculated the phase diagram of a simple model which includes
intersite quadrupolar coupling between uranium ions in the hexagonal lattice.
The model has some of the physics found in the hexagonal uranium-based
materials, like
the in-plane van Vleck magnetism, magnetic character along the c-axis and
degrees of freedom to allow the
possibility of quadrupolar-Kondo effect. We have shown that the intersite
quadrupolar
coupling between uranium ions enhances the critical temperature and plays the
fundamental role in defining the nature of the transitions in the
phase diagram.

	Based on the magnitudes of the specific-heat jumps and the magnetic
moment $\bar{m}$ given by the model around the tricritical point (TP)
when $J_q$ is of order $|J_m|/10$,
we speculate this region as most probable to find the hexagonal uranium-doped
compounds. We are performing a model calculation based on the
Coqblin-Schrieffer \cite{coqblin} formalism to determine whether this
coupling constant ratio can be realized in practice.

	Naturally, the model defined here represents just the first step
in the development of a model for the antiferromagnetic transition in
U-heavy-fermion materials.  We should next introduce explicitly
the conduction electrons, in hence the Kondo effect. The
natural question is what features of the mean-field phase diagram will
survive. The structure of the phase diagram near the tricritical point for
non-zero $J_q$ appears essentially due to the terms
$\delta_{mq} \bar{m}^2 \bar{q}$ and $\epsilon_{mq} \bar{m}^2 \bar{q}^2$
in the free-energy, Eq.~(\ref{eq:free}),
generated by the Hamiltonian (\ref{eq:Hamil}). Gufan {\it et al.}
\cite{gufan} using Landau's mean-field theory have analysed
the phase diagram of a free energy of the form given by (\ref{eq:fexp}),
and they also found the same topology we have around the multicritical points.
Those terms in our model are always
present as long as we have ferroquadrupolar coupling and the hexagonal
symmetry.

	The Kondo effect will not alter the structure of the Ginsburg-Landau
free energy which is dictated purely by symmetry considerations.
However, will renormalize the GL coefficients.  We don't expect the
phase diagram to be strongly modified in the absence of
intersite quadrupolar coupling, since the physics of the magnetic
susceptibility remains that of van Vleck with only  slight modification of the
low temperature values relative to the case of zero
Kondo temperature.   However, once we turn on intersite quadrupolar coupling,
we will
dramatically renormalize the coefficients.   In particular, wherever a $1/T$
appears, in our expressions, we realize that
 at a first pass this is the single-site quadrupolar susceptibility of the U
ions, and given the
quadrupolar Kondo effect, which is two-channel, this will diverge
logarithmically \cite{sacra}.
 Using the single site susceptibility of the
two-channel model as a guide,  we will effect the replacements
 $\delta_{mq} \approx 1/T \rightarrow 1/T_K\ln(T_K/T)$ and
$\epsilon_{mq}\approx 1/T \rightarrow 1/T_K\ln(T_K/T) $ where $T_K$ is the
Kondo temperature.  This will not remove the second-order
quadrupolar and magnetic transitions in our phase diagram, but will renormalize
the
transition temperatures downwards.   As a result of this we anticipate that we
may obtain similar reduced moment values and specific heat jumps to UPt$_3$
with larger intersite quadrupolar
couplings.

In addition to this exploration of how the Kondo effect modifies our already
interesting results, we would like to explore the effects of magnetic field
through magnetostrictive coupling to see
whether we can understand the metamagnetic transition observed in pure UPt$_3$
below 20K in fields of order 20T\cite{franse}.

\acknowledgments

	One of us (V.\ L.\ L.) is greatful to Miloje Makivic for important
discussions in the early stage of this work and to Ross McKenzie for
carefully reading of our manuscript. D. L. Cox gratefully acknowledges
dicussions
with Professors T.\ L.\ Ho and J.\ Wilkins. V.\ L.\ L\'{\i}bero was supported
by
Brazilian Council for Scientific and Technological Development (CNPq),
and by University of S\~ao Paulo (USP). This research was supported by
U.\ S.\ Department of Energy, Office of Basic Energy Sciences, Division
of Materials Research.

\begin{table}
\caption{ Coordinates of the tricritical points (TP) and  critical
end points (CEP) in the phase diagram of the Fig.~\ref{diagram}, for different
quadrupolar coupling $J_q$.}\label{coord}
\begin{tabular}{ccccc}
\multicolumn{1}{c}{$J_q$} &\multicolumn{1}{c}{$t_{TP}$}
&\multicolumn{1}{c}{$-J_m^{TP}$} &\multicolumn{1}{c}{$t_{CEP}$}
&\multicolumn{1}{c}{$-J_m^{CEP}$}\\
\tableline
\ 0.   &0.30 &1.06 &---  &---   \\
\ 0.05 &0.12 &0.64 &0.14 &0.70 \\
\ 0.1  &0.27 &0.79 &0.29 &0.83 \\
\ 0.2  &0.52 &1.01 &0.55 &1.10 \\
\end{tabular}
\end{table}

\begin{table}
\caption{ Coefficients of the expansion of the free energy defined in
Eq.~(\ref{eq:fexp}). Here $d=(2e^{\beta/2} +e^{-\beta/2})^{-1}$.}
\label{fcoef}
\begin{tabular}{l}
	$\alpha_m=-4dJ^2_m\sinh{\beta/2} - J_m$ \\
	$\beta_m =2d^2J_m^4 [(2-\beta)e^{\beta}-5\beta-1-e^{-\beta}]$ \\
	$\gamma_m = 8J_m^6(3\beta^2d^3-4d)\sinh[\beta/2]+
	            8J_m^6\beta d^2(5+e^{\beta})$ \\
	$\alpha_q = 6dJ_q e^{\beta/2} (2+e^{-\beta}-6J_q\beta)$ \\
        $\beta_q  = 108d^2J_q^4\beta^3 (4e^{\beta}-1)$ \\
	$\delta_{mq}   = 12 d J_m^2 J_q (\beta e^{\beta/2} -
			 2\sinh{\beta/2})$ \\
	$\epsilon_{mq}   = 36d^2J_m^2J_q^2( 4\beta e^{\beta/2}-4 e^{\beta/2}
			 -3\beta^2+2\beta +2 +2 e^{-\beta} )$ \\
\end{tabular}
\end{table}

\begin{table}
\caption{ Specific-heat jumps $\Delta C$ of the second-order magnetic
transitions on the phase diagram Fig.~\ref{diagram}, for a few values of
the quadrupolar coupling $J_q$. The magnetic coupling $J_m$ are near the
minimum value to give second-order transitions.}\label{jump}
\begin{tabular}{cccc}
\multicolumn{1}{c}{$J_q$} &\multicolumn{1}{c}{$-J_m$}
&\multicolumn{1}{c}{$\Delta C$} \\
\tableline
\ 0.00&1.10&6.52 \\
\ 0.05&0.60&1.50 \\
\ 0.1&0.68&0.92 \\
\ 0.2&0.85&1.31 \\
\end{tabular}
\end{table}

\figure{ Level structure of the model considered in this work. The doublet
ground-state $|E_{\pm}\rangle=u|\pm4\rangle+v|\mp2\rangle$ is magnetic
coupled to a
singlet excited-state $\sqrt{2}|B\rangle=|3\rangle+|-3\rangle$, which
has energy $\Delta$.
Here $|m\rangle$ is the eigenstate of the z-component of the total angular
momentum $J=4$. The other $D_{6h}$ crystal-field levels are not included,
thus restricting the model to temperatures less than $\Delta$.
\label{3levels}}

\figure{ Mean-field phase diagram of the model defined by the Hamiltonian
Eq.~(\ref{eq:Hamil}). $J_m$ and $J_q$ are the dimensionless magnetic and
quadrupolar coupling between the nearest-neighbouring sites, respectively,
and $t_c \equiv k_BT_c/\Delta$ is the dimensionless critical temperature.
At right of the critical end point (denoted by
CEP), both order parameters have the same critical temperatures. At left, the
transitions are separated into one pure second-order quadrupolar (horizontal
straight line) and one pure first-order magnetic transition (dashed line).
This last one ends at the tricritical point (denoted TP) where a line of
second-order magnetic transition begins and ends at $t=0$ and
$J_m=-0.5(1+3J_q)$. The quadrupolar coupling
enhances the critical temperature $t_c$, increases the absolute
value of the minimum magnetic coupling $J_m$ necessary to have magnetic order,
and reverses the nature of the phase transitions respect to the
tricritical point. \label{diagram}}

\figure{ Temperature dependence of the magnetic moment $\bar{m}$
and quadrupolar moment $\bar{q}$, obtained from Eq.~(\ref{eq:mbar}) and
Eq.~(\ref{eq:qbar}). The temperatures on the x-axis are measured in units
of the crystal-field splitting $\Delta$, and are not in the same scale;
the units of all other quantities are defined in the text.
For $J_q=0$, Fig. (a), (b) and (c) $\bar q$ has the same critical temperatures
as $\bar m$, and behaves like a secondary order-parameter. For $J_q=0.1$,
Fig. (d), (e) and (f) the quadrupolar transitions are separated from
the magnetic ones, except for $|J_m|>|J_M^{CEP}=0.83$. For the same value
of $J_m$ the nature of the phase transitions changes according to
whether $J_q$ is present or not.  No magnetic order is found without
quadrupolar order. \label{behavior}}


\begin{references}

\bibitem{Broholm} C.\ Broholm, H.\ Lin, P.\ T.\ Matthews, T.\ E.\ Mason,
W.\ J.\ L.\ Buyers, M.\ F.\ Collins, A.\ A.\ Menovsky, J.\ A.\ Mydosh,
J.\ K.\ Kjems, Phys.\ Rev.\ B{\bf 43}, 12809 (1991).

\bibitem{Ramirez} A.\ P.\ Ramirez, P.\ Coleman, P.\ Chandra, E.\ Bruck,
A.\ A.\ Menovsky, Z.\ Fisk, E.\ Bucher, Phys.\ Rev.\ Lett.\ {\bf 68},
2680 (1992).

\bibitem{quadruKondo} D.L. Cox, submitted to Sendai Conference on
Strongly Correlated Systems (Sendai-Japan, Sept. 1992), to be published
on Physica {\bf B}.

\bibitem{YUPd} C.\ L.\ Seaman, M.\ B.\ Maple, B.\ W.\ Lee, S.\ Ghamaty,
M.\ S.\ Torikachvili, J.\ S.\ Kang, L.\ Z.\ Liu, J.\ W.\ Allen,
D.\ L.\ Cox, Phys.\ Rev.\ Lett.\ {\bf 67}, 2882 (1991); see also B. Andraka and
A. Tsvelik, Phys. Rev. Lett. {\bf 67}, 2887 (1991).

\bibitem{Geibel} C.\ Geibel et. al., Z.\ Phys.\ {\bf B} Condensed Matter
{\bf 83}, 305 (1991) and {\bf 84}, 1 (1991).

\bibitem{Allen} S.\ J.\ Allen, Phys.\ Rev.\ {\bf 167}, 492 (1968).

\bibitem{sor} K.\ W.\ H.\ Stevens, Proc.\ Phys.\ Soc.\, {\bf A65}, 209 (1952).

\bibitem{nscat} A.\ I.\ Goldman, G.\ Shirane, G.\ Aeppli, B.\ Batlogg and
E.\ Bucher, Phys.\ Rev.\ {\bf B34}, 6564 (1986).

\bibitem{gufan} Yu.\ M.\ Gufan and V.\ I.\ Torgashev, Sov.\ Phys.\
Solid State {\bf 22}, 951 (1980); Yu.\ M.\ Gufan and E.\ S.\ Larin,
Sov.\ Phys.\ Solid State {\bf 22}, 270 (1980).

\bibitem{coqblin} B.\ Coqblin, J.\ R.\ Schieffer, Phys.\ Rev.\ {\bf 185},
847 (1969). See also R.\ Siemann, B.\ Cooper, Phys.\ Rev.\ Lett.\
Phys.\ Rev.\ Lett.\ {\bf 44}, 1015 (1980).

\bibitem{franse} J.\ J.\ M.\ Franse, P.\ H.\ Frings, A.\ de Visser and
A.\ Menovsky, Physica {\bf 126 B}, 116 (1984).

\bibitem{sacra} P.\ D.\ Sacramento, P.\ Schlottmann, Phys.\ Rev.\ B{\bf 43},
13294 (1991).


\end{references}
\end{document}